\documentclass[12pt]{article}
\usepackage{jcappub}
\usepackage{amsmath}
\usepackage{graphicx}
\usepackage{cleveref}

\usepackage{xcolor}
\usepackage[normalem]{ulem}
\usepackage{mathtools} 
\usepackage{multicol} 
\title{The Distribution of Vacua in Random Landscape Potentials}

\author{Low Lerh Feng,}
\author{Shaun Hotchkiss}
\author{and Richard Easther}

\emailAdd{lerh.low@auckland.ac.nz}
\emailAdd{s.hotchkiss@auckland.ac.nz}
\emailAdd{r.easther@auckland.ac.nz}

\affiliation{Department of Physics,\\ University of Auckland, \\Private Bag 92019,\\ Auckland, New Zealand}

\abstract{ Landscape cosmology posits the existence of a convoluted, multidimensional, scalar potential -- the  ``landscape'' -- with vast numbers of metastable minima. Random matrices and random functions in many dimensions provide toy models of the landscape, allowing the exploration of conceptual issues associated with these scenarios. We compute the relative number and slopes of minima as a function of the vacuum energy $\Lambda$ in an $N$-dimensional Gaussian random potential,  quantifying the associated probability density, $p(\Lambda)$. After normalisations $p(\Lambda)$ depends only on the dimensionality $N$ and a single free parameter $\gamma$, which is related to the power spectrum of the random function. For a Gaussian landscape with a Gaussian power spectrum, the fraction of positive minima shrinks super-exponentially with $N$; at $N=100$, $p(\Lambda>0) \approx 10^{-1197}$. Likewise, typical eigenvalues of the Hessian matrices reveal that the flattest approaches to typical minima grow flatter with $N$, while the ratio of the slopes of the two flattest directions grows with $N$. We discuss the implications of these results for both swampland and conventional anthropic constraints on landscape cosmologies. In particular, for parameter values when positive minima are extremely rare, the flattest approaches to minima where $\Lambda \approx 0$ are much flatter than for typical minima, increasingly the viability of quintessence solutions.}
\begin{document}

\maketitle

\section{Introduction}

Over the last two decades cosmology has developed in apparently paradoxical directions. Observationally, the rise of ``precision cosmology'' makes it possible to measure key parameters to within a few percent, setting stringent tests for the detailed evolutionary narrative given by the concordance $\Lambda$CDM cosmology \cite{Planck2018,DES}. Conversely, both slow-roll inflation and string theory, along with the non-zero  dark energy density,  motivates  investigations of multiverse-like scenarios. In particular, stochastic inflation \cite{Linde1986,Adshead2007} suggests that the mechanism which produces astrophysical density perturbations in conventional inflationary models could also support {\em eternal inflation\/}, generating infinite numbers of  {\em pocket universes\/}~\cite{Guth2001}. Likewise, flux-compactified string vacua point to a possible existence of a  {\em landscape\/} \cite{Susskind2003} or {\em discretuum\/} \cite{Bousso2000}   of vacua within the theory. These developments open the door to anthropic explanations of the non-zero vacuum energy density, insofar as a vanishing vacuum energy might have been more plausibly explained by an unknown symmetry, as opposed to the very small but non-zero quantity we have observed. 

While stochastic (or eternal) inflation implies the existence of a multiverse composed of many pocket universes,  this does not require that the ``low energy" (i.e. LHC scale, in this context) physics or vacuum energy differs between pockets:  the naive quadratic inflation model supports eternal inflation, but has a unique vacuum.  In contrast, landscape models have multiple vacua which will, in principle, be populated by stochastic inflation or tunneling. The string landscape, built on the plethora of flux-stablilsed vacua that exist inside Calabi-Yau spaces, is well-known, but is not the unique realisation of this scenario. The complexity of the landscape  and the vast number of vacua it supports is the basis of whatever explanatory power it possesses, and is almost uncountably large (e.g. $10^{500}$ or greater \cite{Douglas2003}), with the inference that essentially almost any value of the vacuum energy can be realised within it. 
 
The detailed properties of any possible landscape are almost entirely unknown.  However, an intriguing approach to landscape scenarios is to strip them down to their barest essence -- by realising multiverse cosmology within a {\em random\/} multidimensional ($N\sim100$ or more) potential of interacting scalar fields.\footnote{Random is used here in the  context of random function theory \cite{GRF1, GRF2, GRF3}.  We refer to random functions rather than random fields, the nomenclature often seen in the mathematical literature, to avoid confusion with the individual scalar fields that are coupled by the potential.} Loosely speaking, a random function of $N$ variables is one whose value at any given point in its range is drawn from some specified distribution. The values of nearby points are typically correlated, ensuring smoothness and continuity. A {\em Gaussian\/} random function is one for which the relevant distribution is a Gaussian, with the common additional stipulation that it is isotropic and invariant under translations. For the purposes of this analysis, a focus on Gaussian random functions is a natural choice. Their structure is highly nontrivial, but their overall statistical properties can be fully described; we will see that these depend only on the dimensionality and a single parameter. 

If we imagine the landscape arising from a superposition of a huge number of individual uncorrelated interaction terms, the Central Limit Theorem suggests that their sum would tend toward a near-Gaussian distribution at any given point. In a ``real'' landscape these terms need not be uncorrelated, but the properties of $N$ dimensional Gaussian random functions can set baseline expectations for landscape cosmologies.  Moreover, the complexity of landscape cosmologies is often seen as a consequence of their dimensionality. However, in many cases the properties of random functions are more sharply defined in the large-$N$ limit than when $N$ is close to unity. Consequently, these scenarios often become better defined as the dimensionality grows.

The first steps in this direction were taken by Aazami \& Easther \cite{Aazami2006} {and Chen et al \cite{Chen2012}}, investigating ensembles of Hessian matrices, which can be viewed as describing extrema in a random landscape. The eigenvalues of the Hessian are all positive at minima in the landscape. For simple random matrix distributions, eigenvalues are likely to be evenly distributed between positive and negative values but fluctuations away from this situation are strongly suppressed at even moderate values of $N$, suggesting that the number of minima is super-exponentially ({$e^{-N^2}$}) smaller than the number of saddles. However, the individual entries of the Hessians of random functions are correlated and therefore not drawn from identical, independent distributions \cite{Battefeld2012,Easther2016}, violating a common premise of simple random matrix theory, rendering minima more probable than this initial analysis suggested.  This overall approach has been  extended in a number of directions \cite{Easther2006, Frazer2011, Henry2009, Marsh2013, Agarwal2011,Yang2012,Masoumi2016,Yamada2018} and this line of inquiry has also motivated studies of the properties of random matrices and random functions at large $N$ \cite{Bray2007,Dean2008,Majumdar2009,Bachlechner2014,Battefeld2012,Fyodorov2013,Masoumi2017}. Similar mathematical problems arise in statistical mechanics, string theory, and complex dynamics \cite{Fyodorov2004,Douglas2004,Douglas2006,Fyodorov2007,Fyodorov2012,Fyodorov2018,Ros2019}.

In the conventional picture of the multiverse, a pocket universe traces out a path in the landscape. Regions of the landscape where the gradient of the potential in ``downhill'' direction(s) is small yield slow-roll inflation. If the landscape is bounded below, typical semi-classical trajectories must terminate at a local minimum, where the value of $V$ fixes the apparent cosmological constant. Moreover, bubbles of space can tunnel between minima, eventually populating all possible minima in the landscape.  The {\em measure problem\/} for the multiverse is notoriously complex, but in this scenario we can use the known properties of Gaussian random functions to estimate $p(\Lambda)$ -- the probability density function of possible cosmological constants -- as a function of the dimensionality of the landscape.  This probability density is of course distinct from the likelihood that an arbitrary observer will measure a given value of $\Lambda$, as this convolves $p(\Lambda)$ with complex anthropic considerations. However, we will show that for a nontrivial volume of parameter space,  $\int_0^\infty  p(\Lambda) {\rm d}\Lambda \ll 10^{-M}$  where $M$ is a number of ${\cal{O}}(100)$, so the probability that the corresponding landscape possesses even one metastable solution with positive vacuum energy can be small. We also find that in scenarios where positive minima are extremely rare, the smallest second derivatives of the potential at these rare minima will be much smaller than those found near typical minima, making these scenarios more likely to support quintessence solutions.

The notion of a landscape which has no positive minima aligns with the \emph{swampland conjecture}, which suggests that all the self-consistent stable minima of a well-behaved theory of quantum gravity might actually be ``underwater'' \cite{Agrawal2018},  located at negative values of $\Lambda$. If true, this would appear to require that the cosmological dark energy was underpinned by dynamical quintessence-like evolution.  The physical basis for the swampland conjecture is entirely separate from our introduction of a Gaussian random landscape, but we will see that the swampland conditions impose a constraint on the parameter space open to these scenarios.  

In what follows we do not work in an analytic, large-$N$ limit. Firstly, we are interested in values of $N$ in the tens or hundreds -- although these are significantly larger than unity, they are not necessarily large enough to ensure that subdominant terms have become unimportant. Rather, we base most of our analysis on direct numerical evaluations of the relevant probabilities.  This has the further advantage of allowing us to pose and answer what might otherwise be complex questions regarding the properties of extrema of these functions, by directly comparing expectations for the eigenvalues of the Hessian matrices at minima in the landscape  
 
 We begin from an $N$-dimensional generalisation of the Kac-Rice formalism \cite{Kac1943,Rice1945} and the  machinery developed and summarised by Bond, Bardeen, Kaiser and Szalay  \cite{BBKS} for studying the statistics of Gaussian random functions in three dimensions. This yields $N$-dimensional integral expressions for the proportion of minima that will have a given value of the potential, $V$. At $N=2$ the full integral can be evaluated analytically; at $N \lesssim 10$ we can directly compute the full integrals numerically; and for $N \lesssim 200$ we evaluate Gaussian approximations to the underlying integrands. The Gaussian approximations are tested against the exact numerical result for $N <10$ where they match closely. These techniques can be generalised to other, more complex questions about trajectories in these potentials, and give information about the ``shapes'' of minima which may inform analyses of tunnelling and inflation. We show that $p(\Lambda)$, the probability density function for the vacuum energy, depends on the dimensionality $N$ and a single free parameter for a Gaussian random landscape, and calculate how $p(\Lambda)$ varies with $N$.

Note that we use both $V$ and $\Lambda$ within the text to refer to vacuum energy density. Explicitly, $V$ is the potential energy density as a function of the underlying fields, $\phi$; and $\Lambda$ refers to the value of $V$ at local minima of the potential.

\section{Random Potentials in $N$ Dimensions}

We treat the potential energy function $V({\phi})$, as a Gaussian random function over an $N$-dimensional field space, $\phi$. We wish to examine minima and saddles, therefore we require the probability density for the value of the potential itself and the values of its derivatives, $\eta_i = \partial V/\partial \phi^i$ and $\zeta_{ij}=\partial^2 V/\partial \phi^i\partial \phi^j$ at individual points in this field space. These variables can be grouped together into a vector ${\bf y} = [V,\eta,\zeta]$, with $\mathcal{N}=1+N+(N^2+N)/2$ independent components. $V$ is a Gaussian random function, therefore the variables in ${\bf y}$ are described by a multivariate Gaussian distribution. A general multivariate Gaussian distribution with $\mathcal{N}$ independent variables has the form
\begin{equation} \label{MultivariateGaussian}
\begin{split}
p({\bf y})d^\mathcal{N}{\bf y} &= \frac{e^{-Q}}{[(2\pi)^\mathcal{N} \mathrm{det}(M)]^{1/2}} d^\mathcal{N}{\bf y} \, ,\\
Q &\equiv \frac{1}{2} \sum_{i,j}^\mathcal{N} \Delta y_i (M^{-1})_{ij} \Delta y_{j} \, .\\
\end{split}
\end{equation}
Here $\Delta y_i$ is the difference between the actual value and the mean value, i.e. $\Delta y_i \equiv y_i - \langle y_i \rangle$, and $M$ is the \emph{covariance matrix}, 
\begin{equation}
M_{ij} \equiv \langle \Delta y_i \Delta y_j \rangle.
\end{equation}
Averages denoted by $\langle \,\,\rangle$ are ensemble averages. We further assume that $\langle V\rangle = \langle \eta\rangle = \langle \zeta\rangle = 0$.\footnote{We can enforce $\langle \eta \rangle = 0$ and $\langle \zeta\rangle = 0$ by assuming $V$ is statistically isotropic in field space. An interesting path for future work would be to examine our results without this assumption of statistical isotropy.} Therefore the probability density only depends on the covariance matrix, $M$, and its inverse.

We next introduce the field space power spectrum, $P$, of the random function $V$, which will be useful for writing $M$ in a concise form. We define it here to be the Fourier transform of the correlation function of $V$, i.e.
\begin{equation}\label{powspec}
\langle V(\phi_1) V(\phi_2) \rangle = \xi(|\phi_1-\phi_2|)= \frac{1}{(2\pi)^N} \int d^Nk e^{i k \cdot (\phi_1-\phi_2)} P(k).
\end{equation}
where $\phi_1$ and $\phi_2$ are two arbitrary points in field space.

We have assumed $V$ is statistically homogeneous and isotropic in field space, therefore the correlation function $\xi$ depends only on $|\phi_1-\phi_2|$ and  $P$ depends only on the magnitude of the Fourier coordinate $k$.\footnote{We work in a field space basis where the field space metric is Euclidean.} The moments of the power spectrum are defined to be
\begin{equation} \label{moments}
\sigma_n^2 = \frac{1}{(2\pi)^N}\int d^Nk (k^{2})^n P(k)
\end{equation}
This gives $\sigma_0^2=\xi(0)=\langle V^2 \rangle$.

We can differentiate equation \eqref{powspec} and then set $\phi_1 = \phi_2$ to get
\begin{align*}
\langle \eta_{i}\eta_{j}\rangle &= \frac{1}{(2\pi)^N} \frac{\partial}{\partial \phi_1^i}\frac{\partial}{\partial \phi_2^j} \int d^Nk e^{i k \cdot (\phi_1-\phi_2)} P(k)\bigg{|}_{\phi_1=\phi_2}\\
&= \frac{1}{(2\pi)^N}\int d^Nk (k_i k_j) P(k)
\end{align*}
The integrand on the RHS is an odd function of both $k_i$ and $k_j$, therefore the integral over all $k$ is zero unless $i=j$. Furthermore, because $k^2 = \sum_i k_i^2$, we know  $\sum_i \langle \eta_{i}\eta_{i}\rangle = \sigma_1^2$. We have assumed the field is statistically isotropic and thus $\langle \eta_{i}\eta_{i}\rangle=\langle \eta_{j}\eta_{j}\rangle$, meaning $\langle \eta_{i}\eta_{i}\rangle=\sigma_1^2/N$ and $\langle \eta_{i}\eta_{j}\rangle=\delta_{ij}\sigma_1^2/N$.


A similar analysis holds for the second derivatives $\zeta_{ij}$, and will yield the rest of the elements of the covariance matrix. In terms of the moments of the power spectrum these are:
\begin{equation} \label{corr}
\begin{split}
\langle VV \rangle &= \sigma_0^2 \\
\langle\eta_i\eta_j\rangle &= \frac{1}{N}\delta_{ij}\sigma_1^2 \\
\langle V\zeta_{ij}\rangle &= -\frac{1}{N}\delta_{ij}\sigma_1^2 \\
\langle\zeta_{ij}\zeta_{kl}\rangle &= \frac{1}{N(N+2)}\sigma_2^2(\delta_{ij}\delta_{kl}+\delta_{il}\delta_{jk}+\delta_{ik}\delta_{jl})
\end{split}
\end{equation}
and all other correlations are zero. For $N=3$ this reduces to equation A1 of \cite{BBKS} (hereafter called BBKS).

This covariance matrix is far from diagonal; we simplify our analysis by using a basis as close to diagonal as possible. The $\eta_i$ variables are already diagonal, as are the $\zeta_{ij}$ terms with $i\neq j$ but the $\zeta_{ii}$ variables are correlated to each other, and also  to $V$. We look for a set of $N$ linear combinations of these $N$ variables that is diagonal, choosing,
\begin{align}
\label{BasisTransform}
x_1 &= -\frac{1}{\sigma_2}\sum_i\zeta_{ii} \nonumber \\
x_n &= -\frac{1}{\sigma_2}\sum_{i=1}^{n-1}\left(\zeta_{ii}-\zeta_{nn}\right),\,\, (2\leq n \leq N) \, .
\end{align}
The $x_n$ here are analogous, but not identical, to BBKS's $x, y, z$. Following BBKS, we also rescale $V$, introducing $\nu = V/\sigma_0$. With this choice of basis the non-zero elements in the covariance matrix become
\begin{eqnarray}
  \langle\nu^2\rangle &=& 1 \nonumber\\
  \langle x_1^2\rangle&=&1 \\
  \langle\nu x_1\rangle &=& \gamma \nonumber\\
  \langle x_n^2 \rangle &=& \frac{2n(n-1)}{N(N+2)},\,\, (2\leq n \leq N) \nonumber
\end{eqnarray}
\noindent where $\gamma = \sigma_1^2/(\sigma_2 \sigma_0)$. The only non-diagonal correlation left is between $\nu$ and $x_1$.

The $Q$ factor in equation \eqref{MultivariateGaussian} becomes
\begin{equation} \label{Q}
\begin{split}
2Q = x_1^2 + \frac{(\nu-\gamma x_1)^2}{1-\gamma^2}+\sum_{n=2}^N\frac{N(N+2)}{2n(n-1)}x_n^2 + \frac{N \pmb{\eta}\cdot \pmb{\eta}}{\sigma_1^2} + \sum_{i,j;i > j}^N\frac{N(N+2)(\zeta_{ij})^2}{\sigma_2^2}
\end{split}
\end{equation}
This is the equivalent of BBKS equation (A4) for $N$-dimensions. Note that the first two terms remain constant for all $N$, but the remaining terms are $N$ dependent.

Equations \eqref{Q} and \eqref{MultivariateGaussian} give, in as diagonal a form as possible, the full probability density function for the function, $V$, and its first two derivatives at an arbitrary point in field space. We are interested in characterising the distribution of \emph{minima} of $V$, and begin by looking at the number density of extrema, i.e. points where $\eta=0$. The total number of extrema in some region in field space is $N_{\rm ext} = \int {\rm d}^N\phi \, \left| {\rm det}\zeta(\phi)\right|\delta^{(N)}[\eta(\phi)]$. The integrand is a delta-function that integrates to give 1 every time a point with $\eta=0$ is passed over in field space. The additional factor $\left| {\rm det \zeta} \right|$ arises because the argument of the delta-function is $\eta$, whereas the integration variable is $\phi$, and $|{\rm det}\zeta|$ is the Jacobian of the transformation between these two variables, i.e. ${\rm d}\eta_i/{\rm d}\phi_j = \zeta_{ij}$. The number \emph{density} of extrema is therefore
\begin{equation}
n_{\rm ext}(\phi) \,\,{\rm d}^N \phi= |{\rm det}\,\, \zeta(\phi)| \delta^{(N)}\left[ \eta(\phi)\right] {\rm d}^N \phi.
\end{equation}

The number density of extrema, $n_{\rm ext}(\phi)$, is itself a random variable. However, we will work with the \emph{expected} number density, which is simply the ensemble average of $n_{\rm ext}$, or
\begin{equation} \label{NumberDensity}
\langle n_{\rm ext}(\phi)  \rangle= \int |{\rm det}\,\, \zeta(\phi)| p(V,\eta=0,\zeta)d\nu d\zeta
\end{equation}
where $p$ is the Gaussian probability distribution given by $\eqref{MultivariateGaussian}$, and for notational convenience we have dropped the ${\rm d}^N \phi$ from each side.

Finally, we need to restrict this integral to count only the extrema that are also local minima. It is easiest to do this by considering the eigenvalues, $\lambda_i$ of the Hessian of the landscape $\zeta$ and restricting them to be positive.\footnote{Note there is an alternative convention where the eigenvalues are defined with the opposite sign, in which case \emph{negative} eigenvalues correspond to minima; e.g. Ref.~\cite{BBKS} (BBKS).}  The Hessian is symmetric by definition. We can also rotate it such that it is diagonal, meaning  $\zeta_{ii}=\lambda_i$, $\zeta_{ij}=0$ and $|{\rm det} \zeta| = \prod_i |\lambda_i|$. The new integration variables are then the eigenvalues themselves and the set of rotation angles required to diagonalise $\zeta$.  In 2D and 3D, the Jacobian of this transformation can be calculated explicitly, which leads to  ${\rm d} \zeta = \prod_{i \neq j}^N |\lambda_i - \lambda_j|\left(\prod_k {\rm d}\lambda_k\right) {\rm d} \Omega$, where ``${\rm d}\Omega$'' represents the $(N^2-N)/2$-dimensional integral measure of the set of rotation angles. The same result holds in $N$ dimensions.\footnote{A proof in $N$ dimensions is given in section V of \cite{Easther2016}, generalising the 3D  BBKS result.} The function $V$ and its derivatives are statistically isotropic. Therefore each set of rotation angles is as probable as any other, and the integral over all the rotation angles must give just a constant. At this point we lose the $\sigma_2$ dependence in the last term in $Q$; $\eta=0$ is the defining property of stationary points, so aside from the dimensionality $N$ itself, the only free parameter explicitly remaining is $\gamma$.

We now have  the ingredients we  need, but  also enforce an ordering on the eigenvalues to avoid dealing with a multimodal integrand.  We define  $\lambda_1 \geq \lambda_2 \geq \lambda_3 \ldots \geq 0$ so our boundary conditions become $x_1\leq x_N\leq x_{N-1} ... \leq x_2 \leq 0$, and the mapping between $\lambda_i$ and $x_i$  follows from Equation~\ref{BasisTransform}. This gives significantly simpler boundary conditions than earlier choices that have been used for diagonalising the covariance matrix, e.g. Ref.\cite{BBKS} (BBKS).

Finally we have the following expression for the expected number density of minima 
\begin{equation} \label{DensityOfPeaks}
\langle n_{{\rm min}} \rangle = A \int_{\lambda_1 \geq \lambda_2 \ldots \geq 0} G \times e^{-Q} \, d\nu \,d^Nx
\end{equation}
\noindent where $G$ has the form\footnote{$G$ can also be expressed in terms of $x_n$ using the inverse transform of Eq. \ref{BasisTransform}.}

\begin{equation}
G = \left(\prod_{i}^{N} \lambda_i \right)\left(\prod_{i<j} |\lambda_i-\lambda_j|\right),
\end{equation} 
$Q$ is given by Eq. \ref{Q} with $\zeta_{ij}=0$ for $i\neq j$ and the constant factor $A$ has absorbed the Jacobian for the transformation between $\lambda_i$ and $x_n$. More generally, we can view the expression
\begin{equation}
{\cal L}(\lambda_1,\cdots, \lambda_N; V, \gamma) = \left(\prod_{i}^{N} \lambda_i \right)\left(\prod_{i<j} |\lambda_i-\lambda_j|\right) e^{-Q}
\end{equation} 
as an unnormalised probability density that gives, for a fixed value of $N$, the (relative) likelihood of finding a minimum at which the landscape has potential value $V$, and the Hessian matrix has normalised eigenvalues $\lambda_i/\sigma_2$ as a function of the single independent variable $\gamma$. This expression is exact;  we immediately see that the prefactor to the exponential enforces a variant of {\em eigenvalue repulsion\/} \cite{Mehta1990}  in that the likelihood of finding a minimum with near-identical $\lambda_i$ is vanishingly small.

\section{Landscape Heuristics}

Our  overall goal is to understand $p(\Lambda)$, the probability density of energy densities (i.e the values of $V(\phi)$) at the minima of a Gaussian random landscape. In particular, we are interested in the proportion of  minima with $V > 0$, which is given by\footnote{We  earlier fixed $\langle V \rangle = 0$. One could consider an ``offset'' landscape, consisting of a Gaussian random function with zero mean, plus a constant, such that $\langle V \rangle \neq 0$ but we do not pursue this here.}
\begin{equation} \label{PminIntegral}
  P(\Lambda >0| N,\gamma) =  \frac{\int^\infty_0 \,\,d\nu \int_{\lambda_1 \geq \lambda_2 \ldots \geq 0} d^Nx \,\, G \times e^{-Q} }{\int^\infty_{-\infty} d\nu \int_{\lambda_1 \geq \lambda_2 \ldots \geq 0} d^Nx\,\,G \times e^{-Q}}
  \end{equation}
 and any specific $p(\Lambda)$ can be computed by dropping the integral over $\nu$ and setting $\nu=\Lambda/\sigma_0$ in the numerator.  
This integral depends on the value of $Q$ (Eq. \ref{Q}) and therefore on the moments of the power spectrum $\sigma_0, \sigma_1$,  and $\sigma_2$, through the single parameter $\gamma$. The allowed range of this variable is $0<\gamma<1$ because $0<\sigma_1^2<\sigma_0\sigma_2$.\footnote{See Appendix \ref{Proof} for a proof.}

\begin{figure}
  \centering
  \includegraphics[width=0.45 \linewidth]{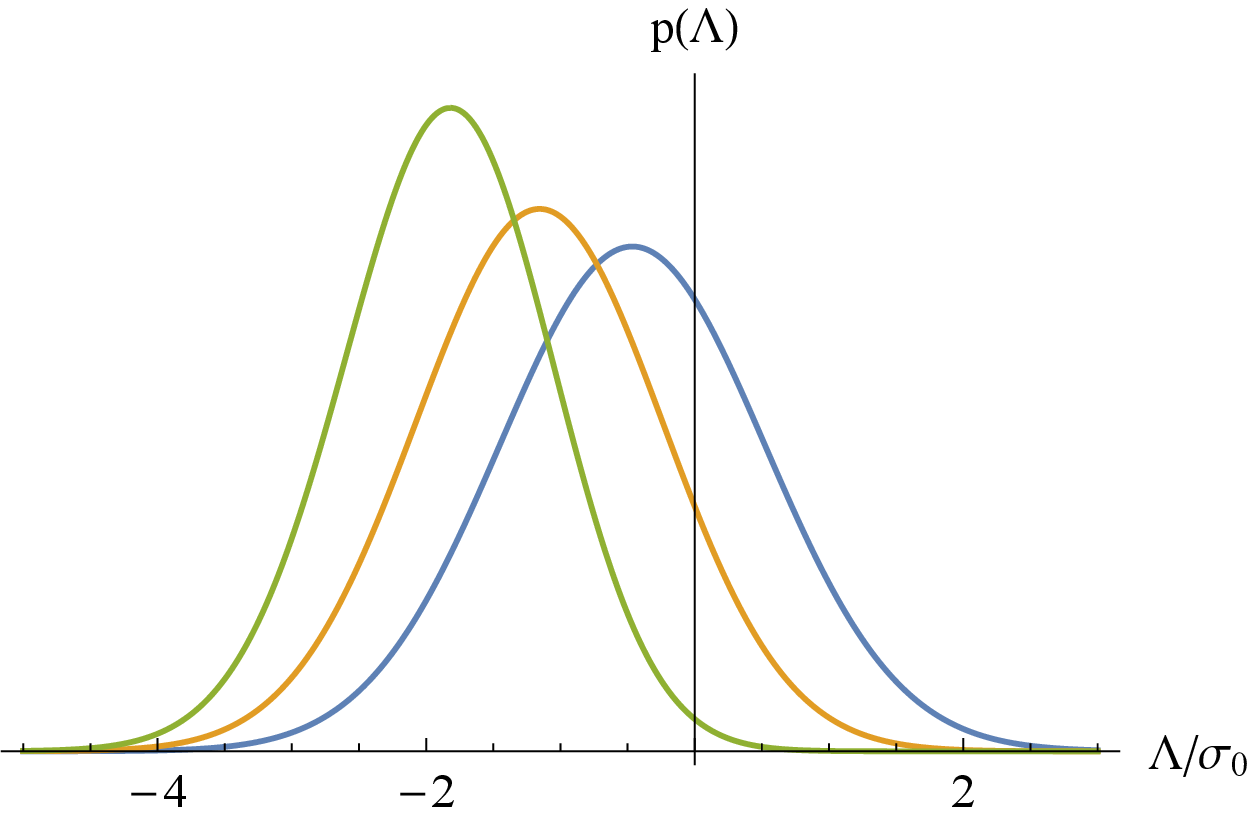}  \hfill
  \includegraphics[width=0.45 \linewidth]{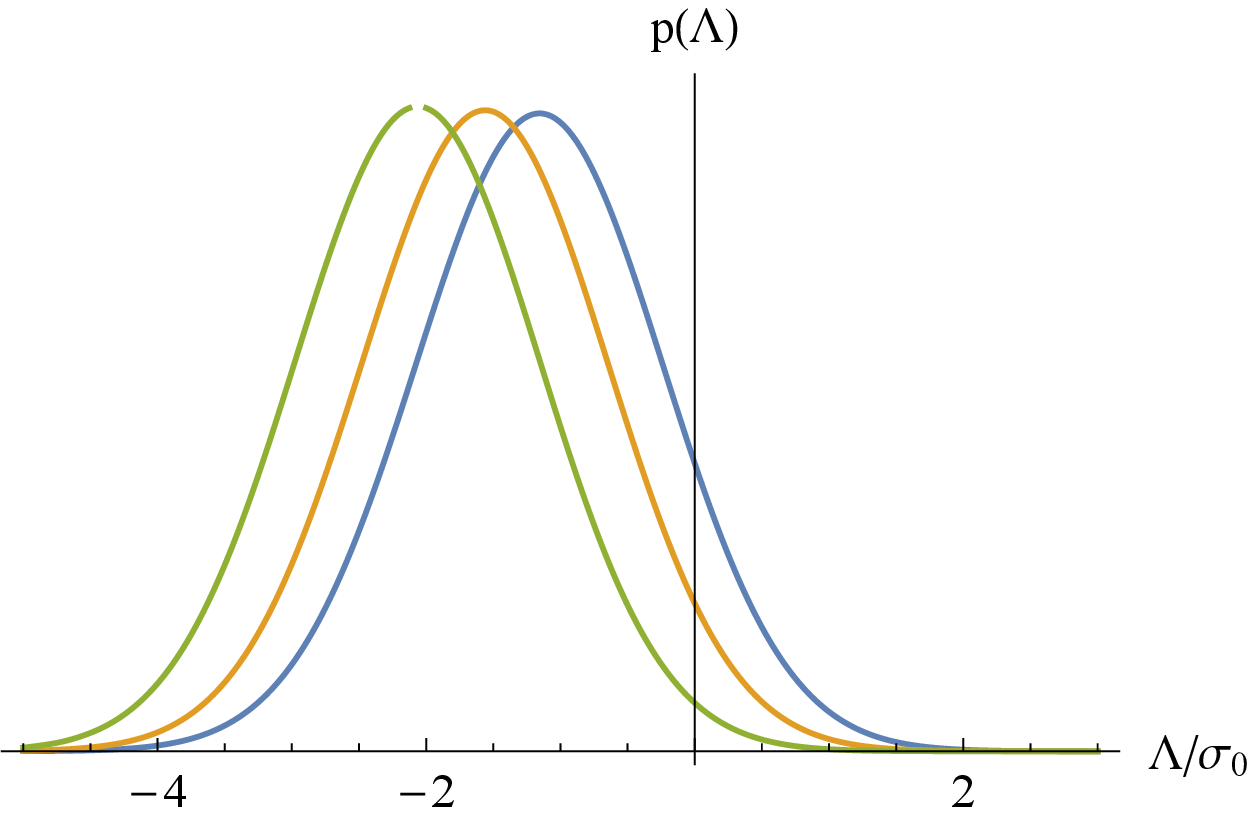}
  \caption{[Left] $p(\Lambda)$ for $\gamma = 0.2$, $0.5$ and $0.8$ and [Right] $p(\Lambda)$ for $N=3$, $5$ and $8$. The distribution moves leftward with increasing $N$ and increasing $\gamma$. The results were obtained by numerically evaluating the full $x_n$ integrals to obtain $p(\Lambda)$.}
  \label{distributions}
  \end{figure}

Broadly speaking, the overall distribution $p(\Lambda | N,\gamma)$ moves to lower values as $N$ and $\gamma$ are increased;  representative results for moderate $N$ are shown in Figure~\ref{distributions}. Moreover, as $\gamma$ increases, the width of the distribution contracts. Consequently, we will find that when $\gamma$ is close to unity and $N \gtrsim {\cal{O}}(10^2)$, $P(\Lambda >0)$ can be almost arbitrarily small, as we quantify below.

This dependence on $\gamma$ can be understood via the second term in $Q$ (equation \eqref{Q}). When $\gamma$ increases, $1-\gamma^2$ decreases, meaning the width of the $\nu$ dependent part of the probability density also decreases. Also, when $\gamma$ increases, the $\gamma x_1$ term grows in magnitude. At minima, $x_1 <0$ and therefore the most probable value for $\nu$ is also negative. As $\gamma$ increases this best fit value becomes more negative. We see both of these effects in the left panel of Figure \ref{distributions}.

\begin{figure}
  \centering
    \includegraphics[width=0.8 \linewidth]{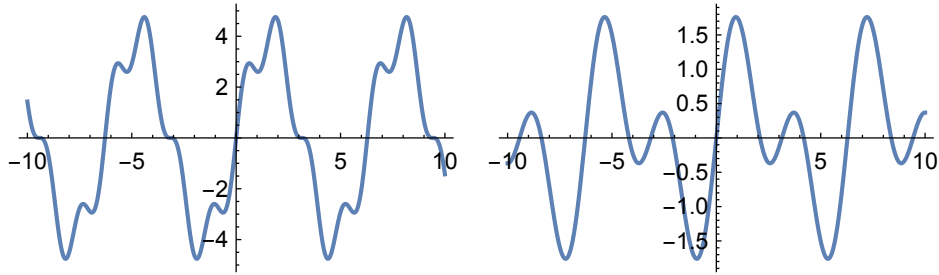}
  \caption{Illustrative realisations of 1D functions. The left figure has a small $\gamma$ and power over a greater range of scales, allowing minima (maxima) to appear in significant numbers above (below) zero. When $\gamma$ is large (right figure) the spectrum is dominated by a smaller range of modes, and most minima are low-lying.}
  \label{examples1}
\end{figure}

One can see this clearly via Figure \ref{examples1}. When $P(k)$ is peaked over a smaller range of scales (i.e. it has a small bandwith) $\gamma$ is closer to one. Similarly, when $P(k)$ is peaked over a wider range of scales, $\gamma$ is closer to zero. This latter case would arise if a potential had  significant ``ripples'' superimposed on top of larger-scale structure in $V(\phi)$. The ripples can produce local minima even when $V(\phi)$ is relatively large, as illustrated in the left hand panel.

Likewise, we can understand the $N$-dependence semi-quantitatively by noting that the number of terms in the polynomial $G$ grows rapidly with $N$, so as $N$ increases, the full probability density at minima prefers larger magnitude eigenvalues, $|\lambda_i|$. Since $x_1$ is the sum of the eigenvalues, as its magnitude grows the $(\nu-\gamma x_1)^2$ term in $Q$ also grows. For minima, $x_1<0$, therefore large $|x_1|$ makes it less likely to find minima with $\nu >0$ (or $\nu <0 $ for maxima). This effect only shifts the mean of the $\nu$ distribution, but does not change its width, which is indeed what we see in the right hand panel of Figure \ref{distributions}. Note that at a generic point in field space $\langle x_1^2\rangle = 1$; however at a minimum $\langle x_1^2 \rangle >1$ due to the effect of $G$.

\begin{figure}
  \centering
  \includegraphics[width=0.8 \linewidth]{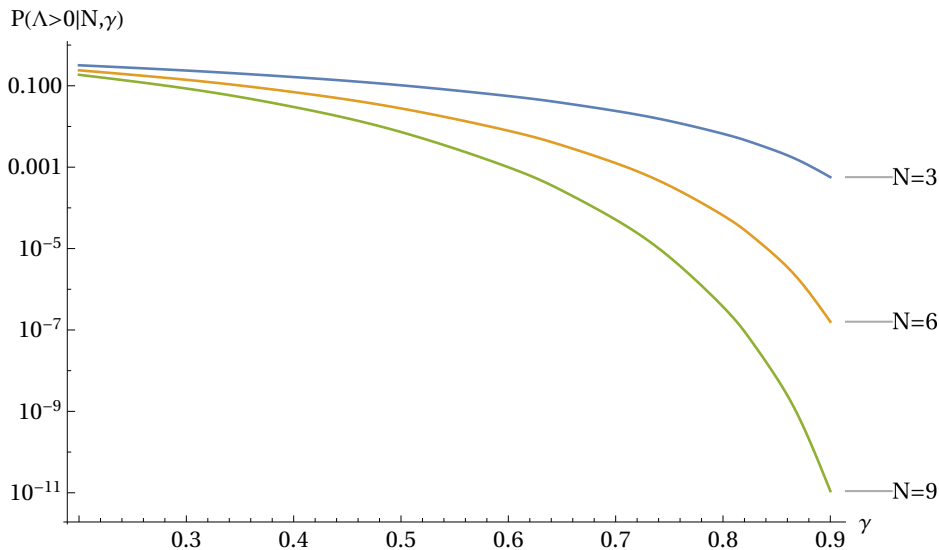}
  \caption{The probability that a given minimum has $V > 0$ as a function of $\gamma$, for $N=3, 6, 9$. We see that both the heuristics are obeyed: the probability decreases with $N$, and  $\gamma$.}
  \label{N6}
  \end{figure}

Given the dependence of $p(\Lambda|N,\gamma)$ on $N$ and $\gamma$ it is clear that as these variables increase $P(\Lambda>0|N,\gamma)$ will involve the increasingly small tail of an increasingly narrow distribution. Figure~\ref{N6} shows $P(\Lambda>0|N,\gamma)$ as a function of $\gamma$ for values of $N$ that are small enough for the relevant integrals to be fully evaluated numerically, and we see that positive minima grow far less likely as $N$ and $\gamma$ increase. 

 For the special case of a Gaussian power spectrum (and not just a Gaussian landscape), 
\begin{equation}
\sigma_n^2 = \frac{2^n \Gamma{\left(n+ \frac{N}{2} \right)}}{\Gamma{\left(\frac{N}{2} \right)}} \frac{U_0^2}{L^{2n}}
\end{equation}
where $U_0$ sets the typical value of $|V|$ and $L$ is a correlation length \cite{Yamada2018}.  Consequently, for this specific power spectrum 
\begin{equation} \label{gammaNgaussian}
\gamma = \sqrt{\frac{N}{2+N}}
\end{equation}
which tends to unity as $N$ becomes large.

\section{Gaussian Landscapes: Evaluating $p(\Lambda)$ and $P(\Lambda>0)$} \label{PeakNumbers}

We now turn to the problem of evaluating $p(\Lambda|N,\gamma)$ and $P(\Lambda>0|N,\gamma)$  for more general parameter values. Many and perhaps all of the following calculations could be performed using analytical techniques in a large-$N$ limit. However, we are  not necessarily working at ``large'' $N$; given the inherently speculative nature of this investigation, there are few  grounds for specifying likely values of $N$;  $N\sim{\cal O}(10^2)$ is not unreasonable, but we cannot necessarily rely on results that may only hold well as $N\rightarrow \infty$. 

Before proceeding further, we reformulate Eq. \ref{PminIntegral} by performing the integrals over $\nu$ analytically. The only dependence of the integrand on $\nu$ is via the first two terms of $Q$, which are independent of $N$. Consequently, we can write the denominator of Eq. \ref{PminIntegral} as 
\begin{eqnarray}\label{denom}
\nonumber {\rm denom} &=&
\int^\infty_{-\infty} d\nu \int_{\lambda_1 \geq \lambda_2 \ldots \geq 0} d^Nx \,\,G \times e^{-Q} \\ 
\nonumber &=&  \int d^Nx\,\, f(x_1,x_2 \ldots )  \int^\infty_{-\infty} d\nu \,\exp\left(-\frac{(\nu- \gamma x_1)^2}{2(1-\gamma^2)}\right)\\
&=& \sqrt{2\pi (1-\gamma^2)}  \int d^N x \,\, f(x_1,x_2 \ldots )
\end{eqnarray}
with
\begin{equation*} f(x_1,x_2 \ldots ) =  G\times \exp\left(-\frac{x_1^2}{2}-\sum_{n=2}^N\frac{N(N+2)}{4n(n-1)}x_n^2 \right).
\end{equation*}
Similarly, the $\nu$ integral in the numerator can be performed to yield an error function
\begin{eqnarray}\label{numer}
\nonumber {\rm numer} &=&
\int^\infty_{0} d\nu \int_{\lambda_1 \geq \lambda_2 \ldots \geq 0}d^N x\,\, G \times e^{-Q} \\ 
&=& \sqrt{\frac{\pi (1-\gamma^2)}{2}}  \int d^N x \,\, f(x_1,x_2 \ldots )\times \left(1+{\rm erf}\left(\frac{\gamma x_1 \sqrt{1-\gamma^2}}{\sqrt{2}}\right) \right)\,\, .
\end{eqnarray}

A third integral we will sometimes wish to evaluate is that for $p(\Lambda)$, which is the (unnormalised) probability density at a fixed $\nu=\Lambda/\sigma_0$
\begin{equation}\label{plam}
p(\Lambda) = \int_{\lambda_1 \geq \lambda_2 \ldots \geq 0}d^N x\,\, G\times \left.e^{-Q}\right|_{\nu=\Lambda/\sigma_0}.
\end{equation}
It is possible to normalise $p(\Lambda)$ by simply dividing by ``denom'' from Eq. \eqref{denom}.

This leaves us with three separate possible $x_n$-dependent integrals  to evaluate: ``denom'' (Eq. \ref{denom}), ``numer'' (Eq. \ref{numer}) and $p(\Lambda)$ (Eq. \ref{plam}).  For $P(\Lambda>0|N,\gamma)$ the final quantity we want to evaluate is ``numer"/``denom".

Up to $N \sim 10$, we can evaluate both ``numer" and ``denom directly. For $N>10$, direct calculation becomes very resource-intensive. To proceed further, we approximate each integrand as a Gaussian. We do this by fixing $\gamma$ and $N$, finding the location and value of the peak of the integrand and then numerically calculating the Hessian of the integrand with respect to the $x_n$ variables around this point. Thus we approximate each integral as
\begin{align}\label{Gauss}
\begin{split}
{\rm integral} &\approx I_p\int_{-\infty}^{\infty} {\rm d}^Nx\,\, \exp\left(-\frac{1}{2}\sum_{l,m}x_lH_{lm}x_m\right) \\
&=I_p \sqrt{\frac{(2\pi)^N}{\mathrm{det} H}}
\end{split}
\end{align}
\noindent where $H_{lm}$ is the integrand's Hessian and $I_P$ is the integrand's value at the peak. $I_P$ and ${\rm det}H$ will of course be different for each integrand, and will depend on $N$ and $\gamma$. For $p(\Lambda)$, $I_P$ and ${\rm det} H$ will also depend on $\Lambda$. It turns out that the numerical location of the peak and its Hessian can be obtained fairly simply for $N\lesssim 200$, for all three integrands, which will be sufficient for our purposes.\footnote{All numerical results in this paper are obtained using Mathematica. We note that the computational efficiency of the calculations can depend strongly on apparently small tweaks in their implementation.}
\begin{figure} 
  \centering
  \includegraphics[width=.45 \linewidth]{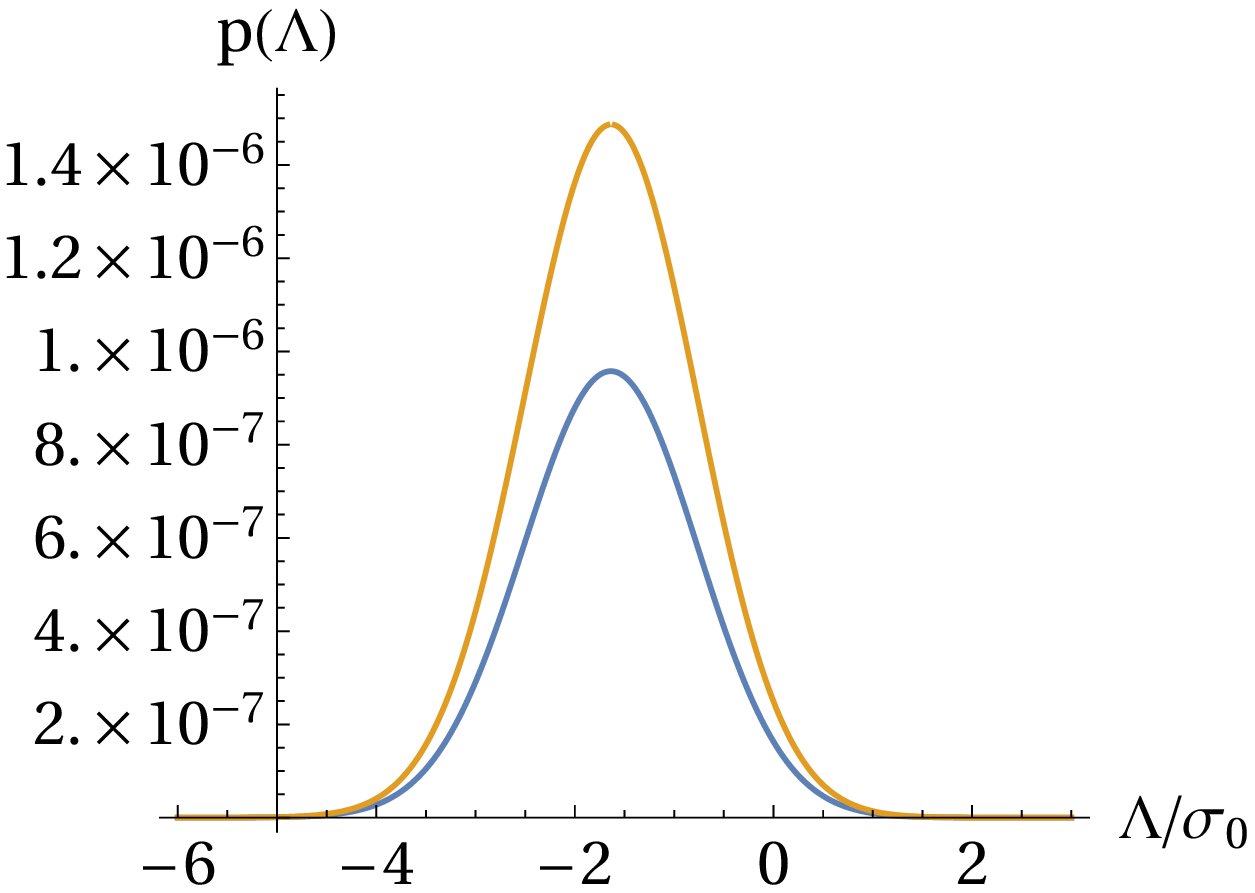} \hfill
   \includegraphics[width=.45 \linewidth]{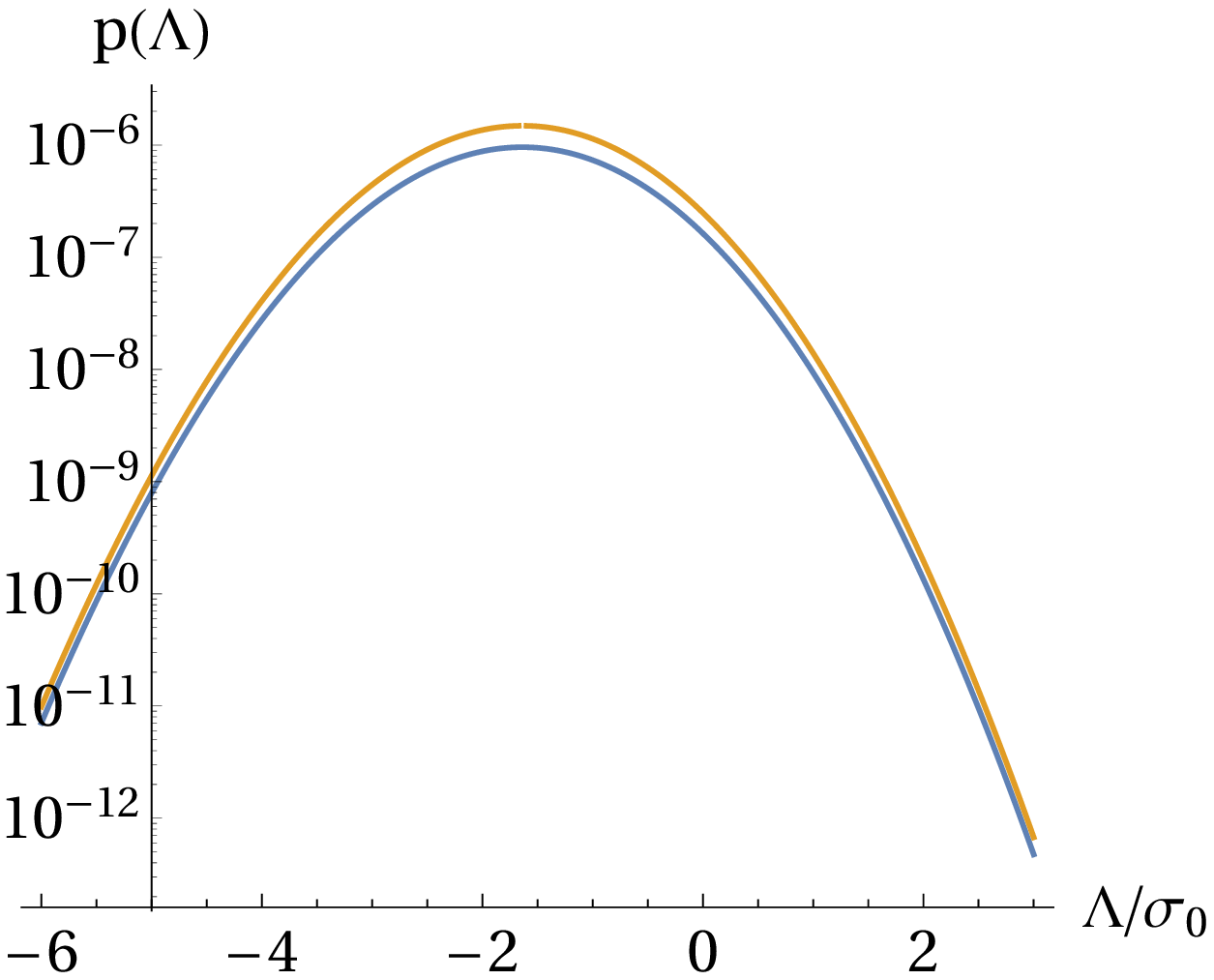} \hfill \\
    \includegraphics[width=.45 \linewidth]{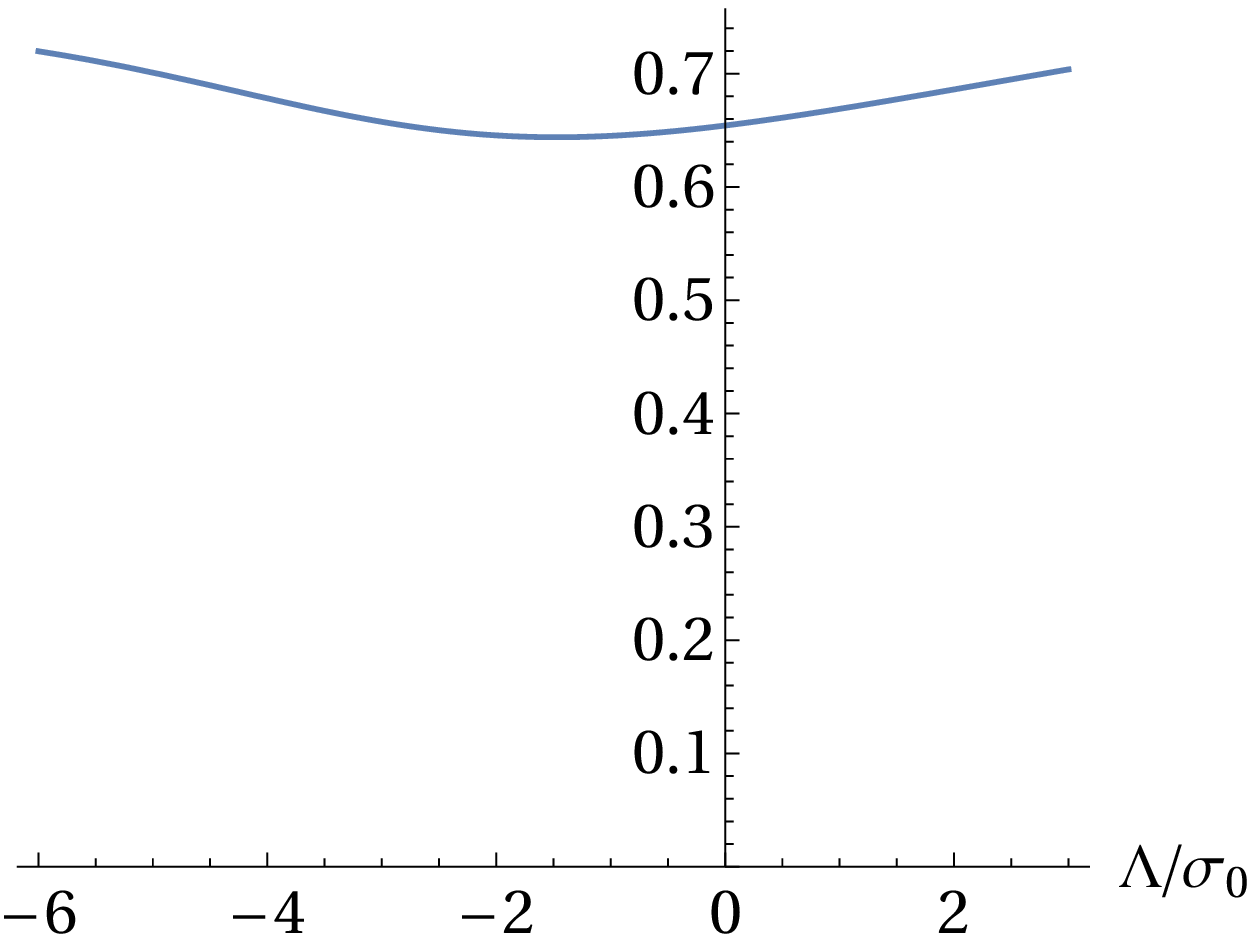}
  \caption{    A comparison of the unnormalised exact $p(\Lambda)$ with the Gaussian approximation, for $\gamma=0.6$ and $N=4$. The left and middle plots show the exact result and the Gaussian approximation, on linear and logarithmic axes; the ratio of the two is show on the right. The Gaussian approximation always  over-estimates the integral, but the ratio is roughly independent of $\Lambda$ and remains close to unity.}
  \label{Comparison}
\end{figure}

\begin{figure} 
  \centering
  \includegraphics[width=0.8 \linewidth]{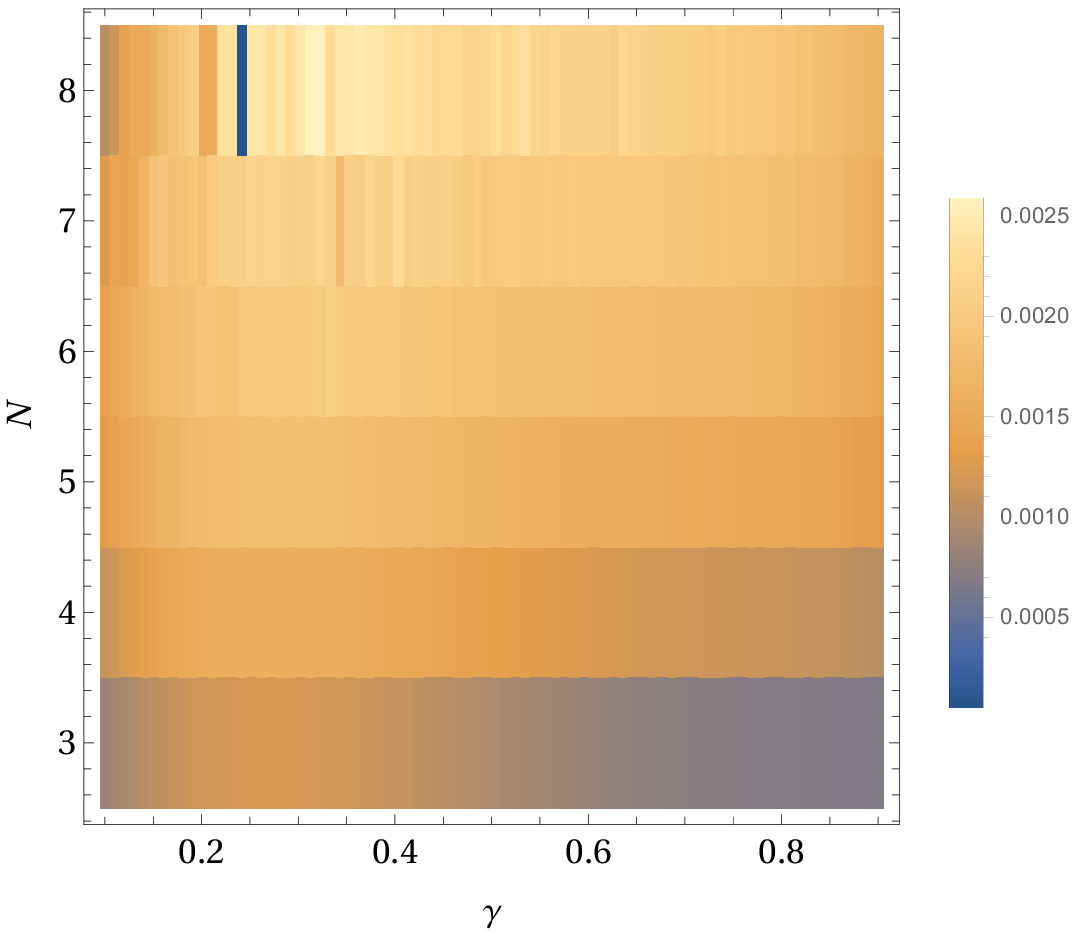}
  \caption{Plot of the relative error in $\rm{log}(P(\Lambda>0|\gamma,N)$ obtained from calculating $P$ using the Gaussian approximation versus the exact numerical result. The error never exceeds $0.3 \%$ over the whole range. For $N>8$ this plot would be dominated by errors in the numerical evaluation of the exact integrals. The error is small everywhere, but see discussion in the text for expectations when extrapolated out to $N\sim  \mathcal{O}(100)$.}

  \label{errcontour}
\end{figure}

A representative comparison between $p(\Lambda)$ from the Gaussian approximation and from the exactly evaluated integral is shown in Figure~\ref{Comparison}. We see that the approximation determines the peak location and shape of $p(\Lambda)$ very well, and determines the amplitude reasonably well, but is off by a small factor. In particular, for this specific $N$ and $\gamma$ the discrepancy between the exact result and the approximation depends very weakly on $\Lambda$, so the normalised results of the Gaussian approximation will be highly accurate ($\sim\%$ level accuracy).

In Figure \ref{errcontour} we plot  the relative error in $\mathrm{log} \left[P(\Lambda>0|N,\gamma)\right]$ between the Gaussian approximation value and the exact numerical value (i.e. $(\rm{log} P_{\rm Gauss})/(\rm{log} P_{\rm exact}) -1$). The plot covers $\gamma \in (0.1,0.9)$ and $N \in (3,8)$.  Within this range, the error in the Gaussian approximation never exceeds $\simeq 0.3\%$. For $N \in (9,12)$ we are still able to evaluate the exact integrals, but not to better than a few $\%$ accuracy. Hence we do not extend Figure \ref{errcontour} to larger $N$ because at $N>8$ the dominant error is in the ``exact'' numerical result, rather than the Gaussian approximation. Nevertheless, over this range, the difference in $\rm{log} P$ between the methods never exceeds the error in the numerical evaluation of the exact integrals.

A similar plot for $P$ would show a maximum error of $\sim 3\%$, and that this error increases slowly with $N$. Fitting this error to an exponential as a function of $N$ and extrapolating to $N\sim 150$ suggests that the Gaussian approximation would underestimate the normalised probability by a factor of $\lesssim 3$ for the worst fitting value of $\gamma$. However, our analysis focuses on the logarithm of the probability; in what follows scenarios with (for example) $P(\Lambda>0) \sim 10^{-500}$ and $\sim 10^{-498}$ are functionally equivalent, despite being different by a factor of 100. The relative error in $\rm{log} P$ is the key quantity and that remains tiny.

These results justify our use of the Gaussian approximation over the full range of $\gamma$ and at larger values of $N$ than those for which we can evaluate the exact numerical integrals. 

\begin{figure} 
  \centering
  \includegraphics[width=.6 \linewidth]{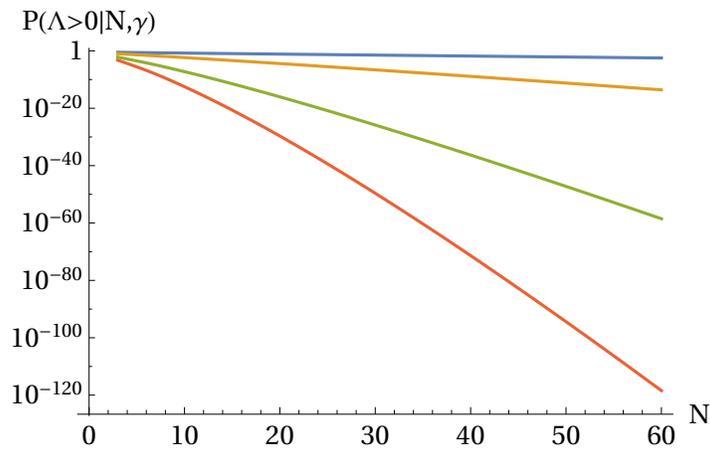}
  \caption{The probability that a given minimum has $\Lambda>0$, as a function of $N$ for $\gamma=0.2$, $0.5$, $0.8$ and $0.9$ (top to bottom) calculated using the Gaussian approximation.}
  \label{PVaryingWithNGaussian}
\end{figure}
  
In  Figure~\ref{PVaryingWithNGaussian} we show $P(\Lambda>0|N,\gamma)$, with a logarithmic $y$-axis, as a function of $N$ for several values of $\gamma$. We use the Gaussian approximation for the integrals over the $x_i$. We see that for sufficiently large $N$ the lines are roughly linear with $N$, but the constant of proportionality incre
ses rapidly as $\gamma$ approaches unity; for $N=30$ and $\gamma = 0.9$ only 1 in $10^{50}$ minima in the landscape will have a positive cosmological constant.

\begin{figure} 
  \centering
  \includegraphics[width=0.8 \linewidth]{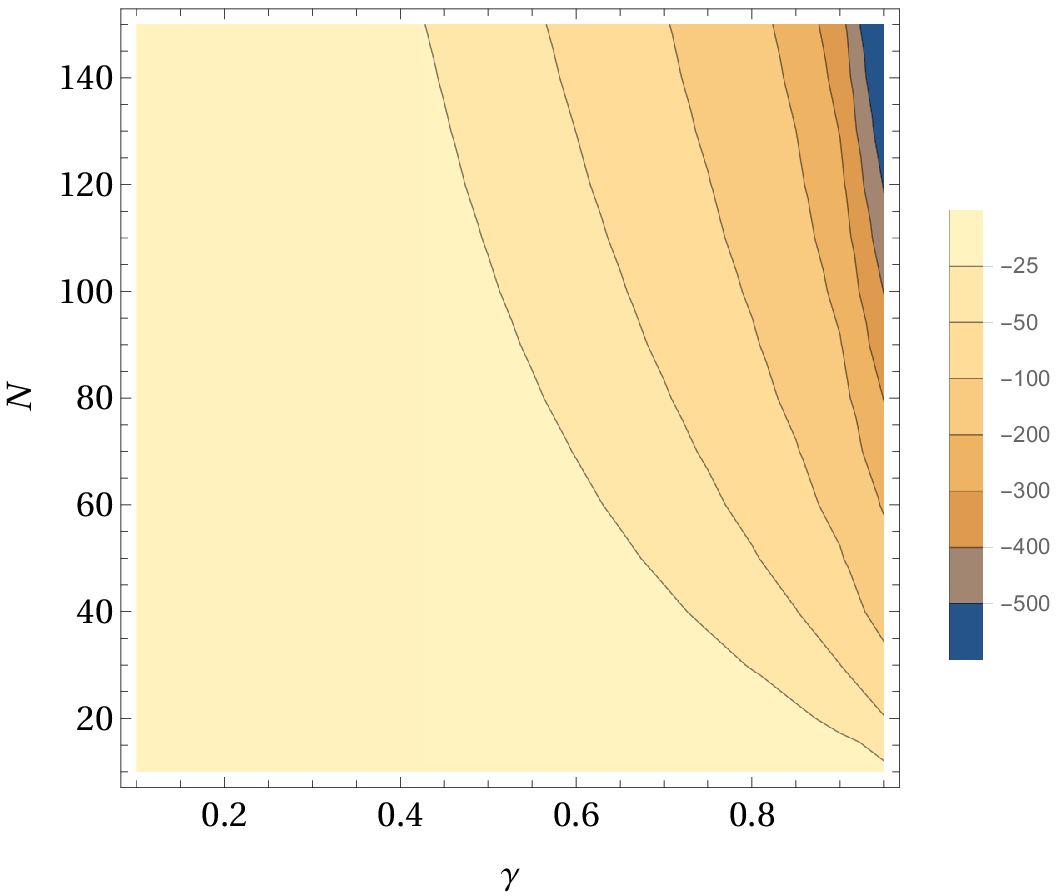}
  \caption{We plot $\log_{10}(P(\Lambda>0|N,\gamma))$ as a function of $N$ and $\gamma$; as $N$ becomes moderately large, less than 1 in $10^{-500}$ minima have $\Lambda>0$, for $\gamma$  close to unity.}
  \label{fullcontourplot}
\end{figure}

 Figure~\ref{fullcontourplot} displays  $P(\Lambda >0 |N,\gamma)$ as a function of $N$ and $\gamma$. For sufficiently large $N$ and $\gamma$, there are fewer than 1 in $10^{500}$ minima with $\Lambda>0$ in the whole field volume. While there is absolutely no guidance as to the expected number of minima in any landscape, $10^{500}$ is a rough benchmark and for  reasonable values of $N$ a Gaussian random landscape could easily contain no {\em positive\/} minima and a very large number of negative minima.  The minimal physical interpretation of this result is that it provides a counterexample to the widespread and implicit assumption that a potential with many minima  necessarily supports an anthropic solution to the cosmological constant problem; even with a ``landscape'' it is still necessary to consider $P(\Lambda>0)$.

\section{Eigenvalue Distributions and Slopes}

In order to compute  Gaussian  approximations to the integrals in Equation~\ref{PminIntegral} we must locate the position of the peak as a function of the $x_i$. However, this calculation yields much more information than just $p(\Lambda  |N,\gamma)$, as the $x_i$ are a linear combination of the $\lambda_i$, the eigenvalues of the Hessian matrix of $V(\phi)$ at the minima. In particular, because we work with the full probability density function rather than an approximation, we can compute the likely shape of minima at values of $V(\phi)$ far from the peak of $p(\Lambda)$ as easily as we can for typical values.

Before we go into the details, we note two things. First, the distribution of eigenvalues at the most likely $\Lambda$, for a given $N$, is independent of $\gamma$. This can be seen by looking at the overall probability density; it is maximised with respect to $\nu$ when $\nu = \gamma x_1$; enforcing this eliminates $\gamma$ from the full  probability density. Second, the eigenvalues themselves are linear combinations of $\sigma_2x_n$. Recall that the eigenvalues, $\lambda_i$, are eigenvalues of the Hessian, $\zeta_{ij}$ and the scale of $\zeta_{ij}$ is set via $\sigma_2$, see Eq. \eqref{corr}. Therefore in the figures in this section we plot the rescaled eigenvalues $\lambda_i/\sigma_2$ (which we nonetheless refer to as `eigenvalues' out of convenience).

\begin{figure} 
  \centering
  \includegraphics[width=.9\linewidth]{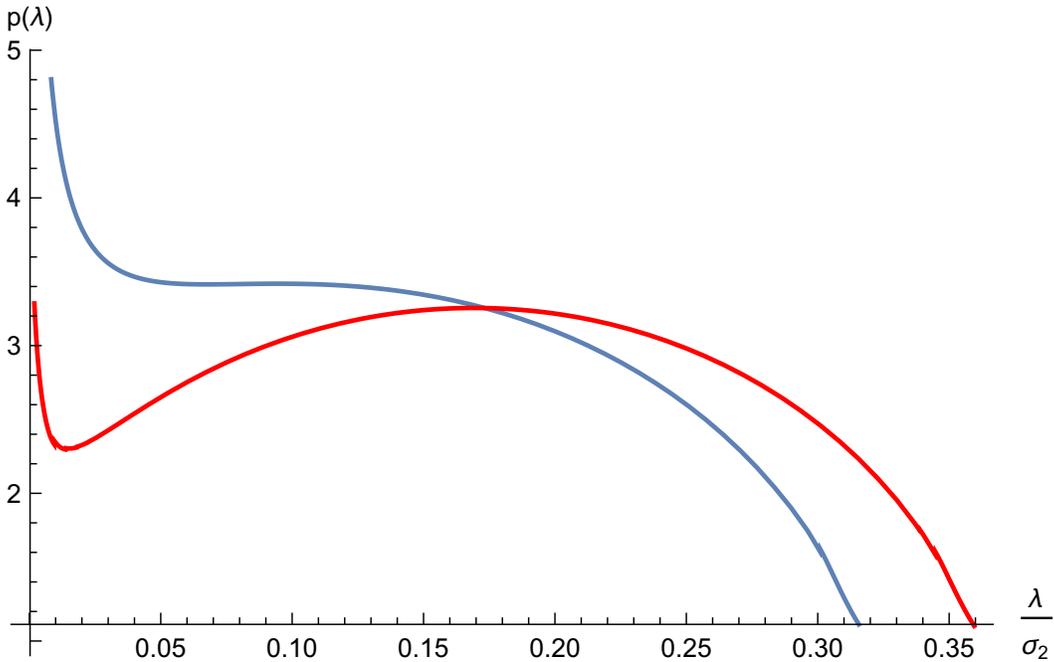}
  \caption{The expected distribution of eigenvalues at the overall peak (numerically calculated to have $\Lambda/\sigma_0 = 15.64$, red line) and at $\Lambda=0$ (blue line) for $N=100, \gamma = 0.9$. The horizontal axis corresponds to the eigenvalues at that point, while the vertical axis is the expected number density of each eigenvalue. Note the distribution at the overall peak is independent of $\gamma$ (see text). Compare Figures 8 and 9 in Yamada and Vilenkin (\cite{Yamada2018}).}
  \label{eigendist}
\end{figure}

In Figure \ref{eigendist} we show, for $N=100$, the expected distribution, $\rho$, of the eigenvalues at minima for $N=100$ and $\gamma=0.9$. We plot this for both the most likely value of $\Lambda$ and for $\Lambda = 0$. The eigenvalues are all correlated. Therefore, calculating this in full detail would require a large set of $N$, $N$-dimensional integrals, one for each eigenvalue. Instead we approximate $\rho$ by the distribution of eigenvalues at the most probable point in the probability density function. Unfortunately, this gives us only $N$ eigenvalues. If we were to bin these eigenvalues and plot a histogram the resolution would be very poor. Instead, we order the eigenvalues from smallest to largest as $\lambda_n$. If we invert this it gives us, for each value of $n$, the value of $\lambda$ for which there are $n$ eigenvalues equal or smaller, i.e. $n(\lambda)$. Other than at $\lambda\simeq 0$ this function is smooth. Therefore we can smoothly interpolate $n(\lambda)$ to non-integer values of $n$. We thus finally obtain $\rho(\lambda)$ by differentiating $n(\lambda)$ with respect to $\lambda$, i.e. $\rho(\lambda) = {\rm d}n/{\rm d}\lambda$. We expect this to be a good approximation of the true $\rho(\lambda)$ when the true $\rho(\lambda)$ is non-negligible and has no sharp features, which is true everywhere except near $\lambda=0$.

For the most likely value of $\Lambda$, we see something similar to the Wigner semicircle (which is the limiting eigenvalue distribution without the constraint of a minimum \cite{Wigner1955,Wigner1958}), with a small upturn at the lowest-lying eigenvalues. For the distribution at $\Lambda=0$ we see a much more significant upturn. Intuitively, this makes sense: for $\Lambda > 0$, minima are rare, so we would expect that they will become increasingly ``marginal'' and the eigenvalues of the Hessian to take on smaller values.

The distribution of eigenvalues in Figure \ref{eigendist} is very similar to that found by Yamada and Vilenkin \cite{Yamada2018}, except we do not see the sharp downturn in probability for very small eigenvalues. This is because we are only sampling a single point in the full probability density function to estimate the distribution of eigenvalues at a minimum. To resolve the downturn, we would need significantly more eigenvalues. Unfortunately we cannot simply go to larger $N$ because the location of the downturn decreases with $N$. Therefore we will never be able to resolve the downturn using this method. However, because the integrand in Eq. \ref{DensityOfPeaks} vanishes for $\lambda_N=0$, if we had sufficient resolution, we must see a vanishing probability at that point. We verified that, if we fix all the eigenvalues except the smallest to their most-likely values, and allow only the smallest eigenvalue to vary, $\lambda_N=0$ is indeed excluded.

\begin{figure} 
  \centering
  \includegraphics[width=.9\linewidth]{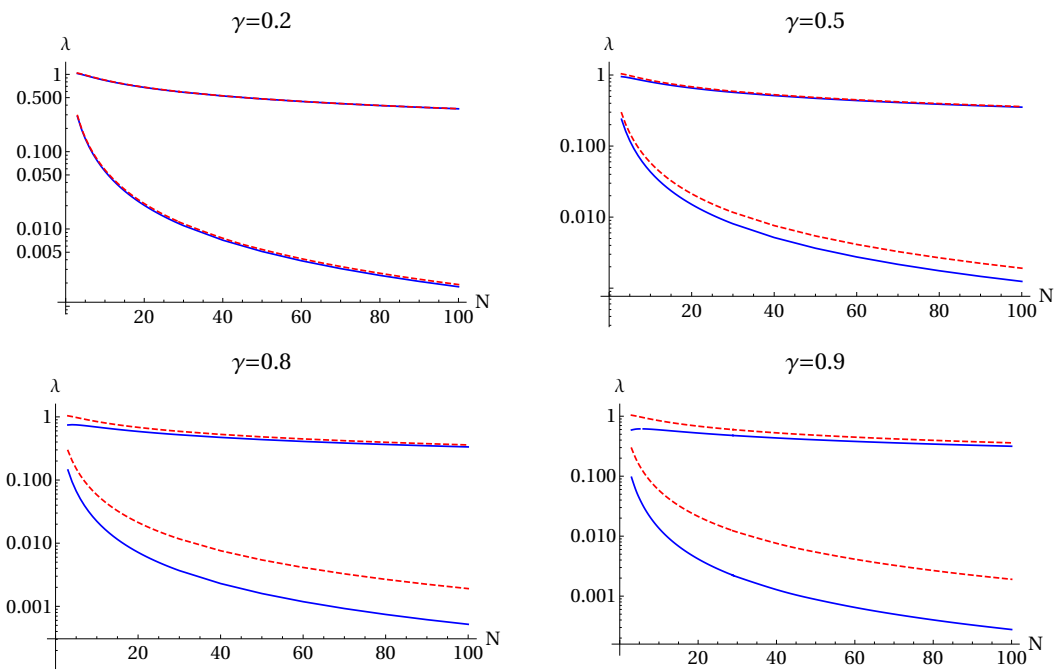}
  \caption{We show the expected minimum and maximum eigenvalues of the Hessian at a metastable vacuum in the landscape, as a function of $N$ for four values of $\gamma$.  Dashed lines denote eigenvalues at the most likely value of $\Lambda$; solid lines indicate values at $\Lambda=0$. The expected values of the smallest eigenvalue decreases as $N$ increases, while the maximum value is roughly unchanged. }
  \label{eigen}
\end{figure}

In Figure \ref{eigen} we plot the distribution of the biggest and smallest rescaled eigenvalues for different values of $\gamma$ and $N$ at two points: at the most likely value of $\Lambda$, and at $\Lambda=0$.\footnote{Although the distribution of eigenvalues at the most likely value of $\Lambda$ is independent of $\gamma$, once we fix $\Lambda=0$, it is no longer independent of $\gamma$.} We see that in all cases the largest eigenvalue is a little less than unity. Conversely, as $N$ increases the smallest eigenvalue approaches zero. When $\gamma$ approaches unity, almost all of the minima in the landscape  occur at negative values of $\Lambda$ and the relative handful of minima above the waterline become even more ``elongated'' as the lowest-lying eigenvalue becomes even closer to zero. This will have implications for quintessence scenarios in any landscape, as it points to a mechanism for producing minima which are almost guaranteed to have a handful of very gradual approaches, as we discuss in the following Section. We further note that the order-of-magnitude estimate by Yamada and Vilenkin \cite{Yamada2018}, that the smallest eigenvalue is of order $\frac{1}{N}$, is accurate.
 
\begin{figure} 
  \centering
  \includegraphics[width=.45\linewidth]{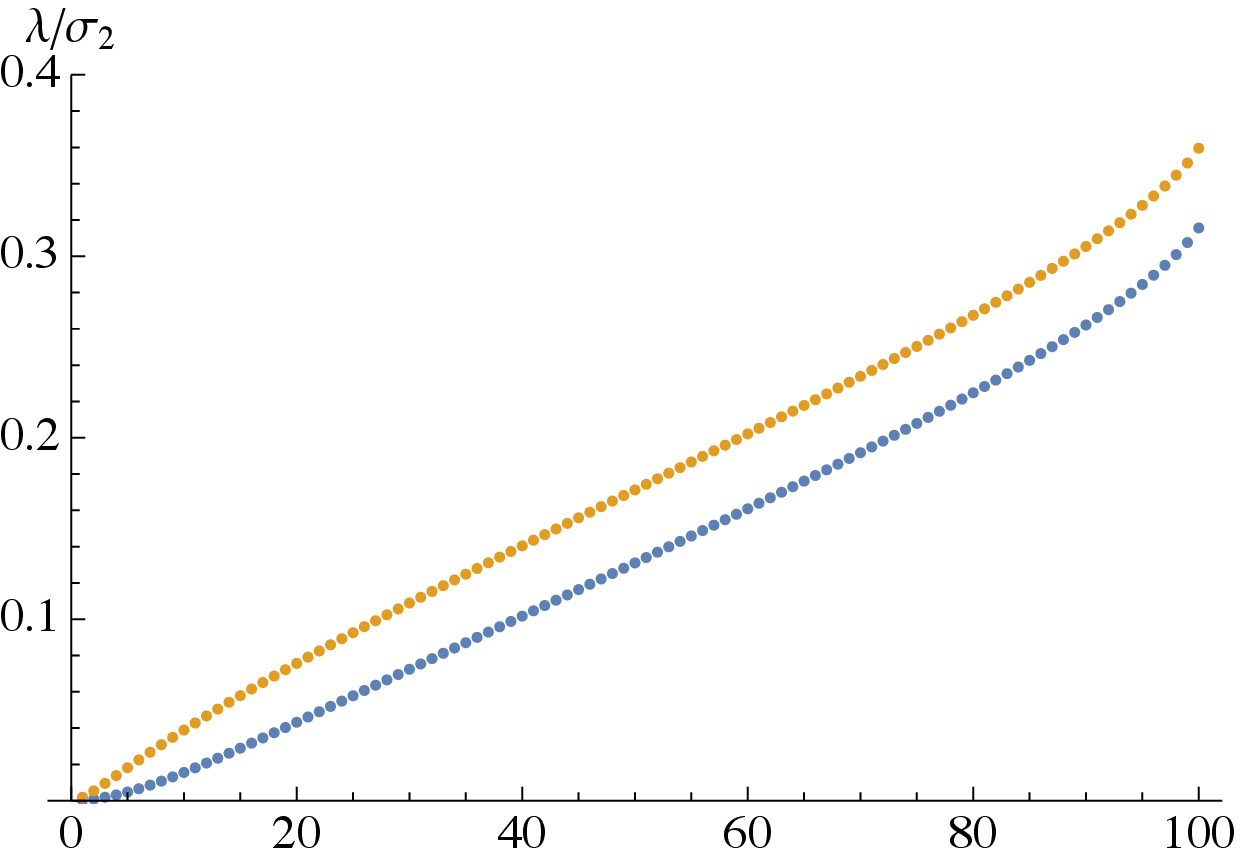} \hfill \includegraphics[width=.45\linewidth]{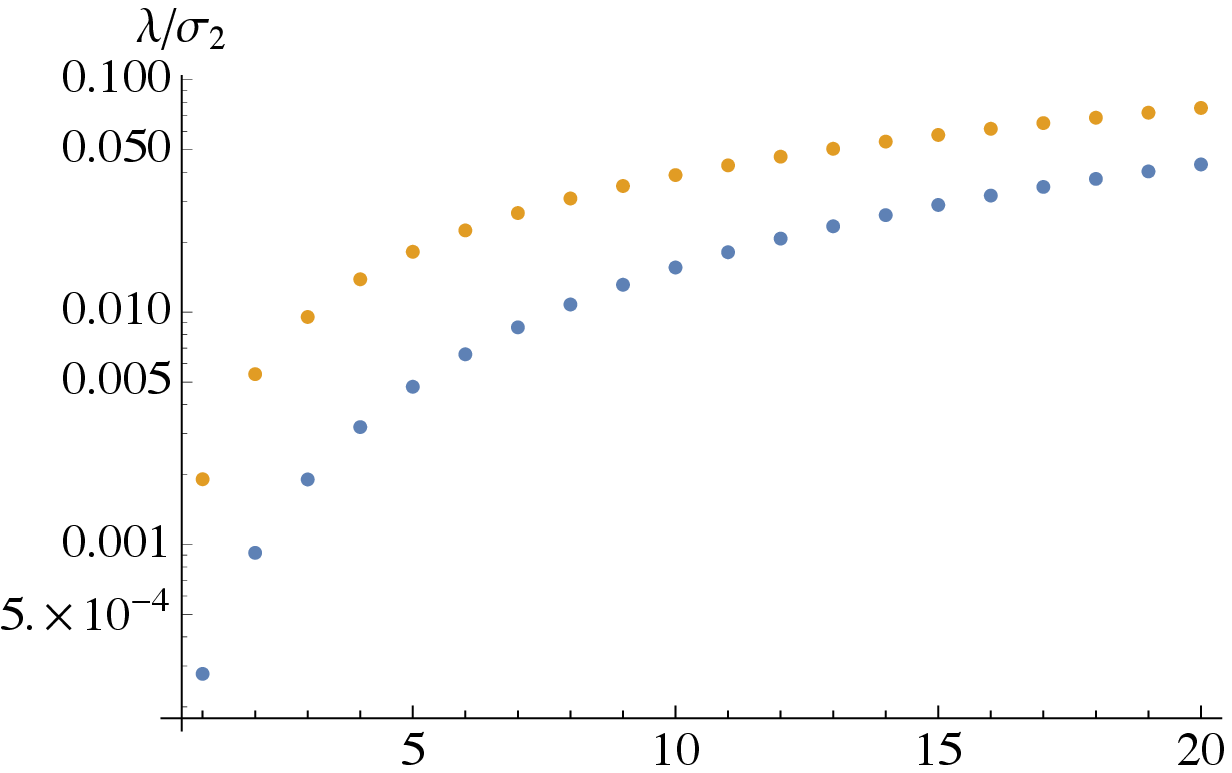}
  \caption{The expected eigenvalues are plotted for $\gamma=0.9$ and $N=100$; the left hand plot shows the most likely value of each of the 100 $\lambda_i$; the right hand plot shows the smallest eigenvalues on a log scale. The upper points correspond to the value for $\Lambda$ for which $p(\Lambda| N,\gamma)$  is peaked; the lower points correspond to $\Lambda=0$.}
  \label{eigendist2}
\end{figure}
 
\begin{figure} 
  \centering
  \includegraphics[width=.6\linewidth]{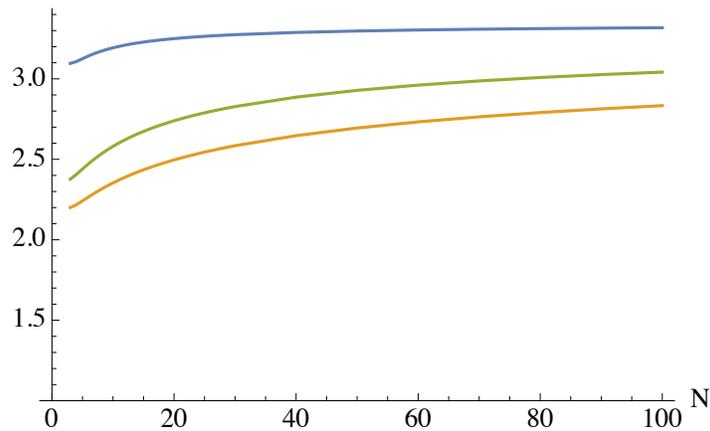}  
  \caption{The expected ratio of the two smallest eigenvalues is plotted as a function of $N$.   From bottom to top the lines are i) the ratio at the peak  of the distribution (which does not depend on $\gamma$), and $\Lambda=0$ for $\gamma =0.5$ and $0.9$.}
  \label{ratio}
\end{figure}

In Figure \ref{eigendist2} we show the most likely value for each of the $N$ ordered eigenvalues. We also zoom in on the smallest eigenvalues in the right hand panel. We see that the eigenvalues are consistently smaller for the marginal, $\Lambda=0$ minima. 
Conversely, as we show in Figure~\ref{ratio} the ratio between the two lowest lying eigenvalues at a minimum increases slowly with $N$ and with $\gamma$; interestingly, not only does the value of the lowest lying eigenvalue decrease as  $N$ increases, the ratio between this and the next-smallest eigenvalue increases with $\gamma$, which will have implications for cosmological models that are based on dynamical scalar fields in a landscape.

\section{Discussion and Implications for Multiverse Cosmology}

 We have  written down an explicit and tractable probability density function for the extrema of an $N$-dimensional  Gaussian random function. We have presented results for the statistics of minima as a function of the dimensionality $N$ and a single additional parameter $\gamma=\sigma_1^2/(\sigma_0\sigma_2)$, which is determined by the first three moments of the function. The results confirm the intuitive expectation that the probability a given minimum will have $\Lambda > 0$ decrease as $N$ and $\gamma$ increase. We calculate precise values for this probability for $N \lesssim10$ and find that it is well-approximated by a Gaussian integral. Directly evaluating the Gaussian approximations out to $N \gtrsim 100$ shows that $P(\Lambda > 0| N, \gamma) \lesssim 10^{-500}$ for a nontrivial segment of parameter space.  

For the specific case of a Gaussian random function with a Gaussian power spectrum, Eq.~\ref{gammaNgaussian}, $\gamma \approx 1 -  1/N$ when $N$ is large.  Consequently, $P(\Lambda>0 | N)$ decreases super-exponentially with $N$; beyond the dependence of $P(\Lambda>0 | N,\gamma)$ on $N$ at fixed $\gamma$,  $\gamma(N)$ increases toward unity, further decreasing $P(\Lambda>0$); if $N=100$, $\gamma \approx 0.99$ $P(\Lambda>0) \approx 10^{-1197}$. For a generic landscape $N$ and $\gamma$ need not be correlated, but for this specific scenario the expected number of positive maxima will drop precipitously as $N$ increases.

Given a huge number of individual, uncorrelated interaction terms, the Central Limit Theorem suggests that their sum would tend toward a near-Gaussian distribution at any given point. However, we should be clear that we are not proposing an $N$-dimensional Gaussian random function as a concrete realisation of the landscape hypothesis; rather, the purpose of this work is to use Gaussian random functions to explore the possibilities opened by landscape scenarios. In particular, many of the situations we consider involve the extreme tails of distributions which, in practice, are often divergent from the predictions of the underlying distribution.

The above results provides a concrete example of a scenario demonstrating that simply having a vast number of minima does not guarantee any of them will be ``above the waterline''.  Stochastic and anthropic solutions to the apparent tuning problem faced by a na\"ive cosmological constant tacitly assume that a vast number of minima can be found with $\Lambda>0$ but the calculation here makes it clear that this is far from guaranteed. Given an infinitely large landscape (in terms of its volume in field-space), any finite dimensional Gaussian random landscape will have an infinite number of minima with $\Lambda>0$, since $P(\Lambda >0)$ is always non-zero. However, typical landscape scenarios posit that the total number of minima is very large but still finite; in these cases it is possible that the number of minima with $\Lambda >0$ will be zero.  

Interestingly, the ``swampland'' hypothesis posits that $M_{Pl} |\nabla_\phi V|/ V \gtrsim c$ where $c$ is a number of order unity and $M_{Pl}$ is the Planck mass \cite{Agrawal2018,Ooguri:2018wrx}; that is, the landscape potential has no minima with  positive vacuum energy. The swampland consists of apparently viable effective quantum field theories coupled to gravity which would in fact be inconsistent with a full quantum theory of gravity. This hypothesis can be construed as a constraint on the parameter space of a Gaussian random landscape. Given a Gaussian random landscape where the volume of field space accommodates an expected $10^{M}$ minima, with $M\gtrsim 100$, this immediately translates into a nontrivial restriction on $N$ and $\gamma$ which could be imposed from  Figure~\ref{fullcontourplot}. 

The $\gamma$  parameter is a measure of the ``bandwidth'' of the landscape function; a Gaussian random function with a power spectrum that has support over a larger range of $k$ values will have a smaller value of $\gamma$. We do not draw strong physical conclusions from this observation, but it is worth noting that a Gaussian random landscape that is consistent with the swampland hypothesis is likely to have contributions from a narrower range of scales in field space.  
 
 It is a common (and sometimes implicit) assumption that the typical value of the vacuum energy density in the multiverse is  fixed by a UV threshold, most typically the Planck scale. This would mean that the energy density $V$ should have a distribution with width, $\sigma_0 \sim 1$, in Planck units. In Planck units the present-day dark energy density is $\sim 10^{-122}$. Since $p(\Lambda)$  has a width $\simeq \sigma_0 (1-\gamma^2)\sim{\cal{O}}(1)$, the distribution will be effectively uniform between $\Lambda=0$ and very small values; e.g. $\lesssim 10^{-100}$. Consequently, while the anthropically allowed range of $\Lambda$ is ill-defined, it is certainly much narrower than the typical width of this distribution.\footnote{E.g. recombination occurs at a redshift $z\sim 1000$ when the density was $\sim 10^9$ times higher than at present. However, a range of $\Lambda$  $10^{10}$ larger than the present dark energy density is effectively infinitesimal compared to the total range.} Consequently, the global form of $p(\Lambda)$ does not provide a meaningful preference for lower values of $\Lambda$ within the anthropically allowed range. 

The discussion above implicitly assumes that the dark energy density is fixed, and field velocities are zero. However, quintessence \cite{Zlatev:1998tr,Tsujikawa2013} has been widely considered, and may be required if the swampland conjecture is valid \cite{Marsh:2018kub,Kinney:2018kew,Heisenberg:2018yae}. Interestingly, we see that as $N$ increases the expected values of the lowest lying eigenvalues approach zero, and that this effect is stronger if $\gamma$ is such that $P(\Lambda>0)$ is vanishingly small. This will be analysed in more detail in future work, but it holds the prospect that Gaussian random landscapes would naturally produce minima with very small positive $\Lambda$  for which the second derivative of $V$ was very small in one direction. As discussed in the previous section, the ratio of the first and second smallest eigenvalues of the Hessian of $V$ at a minima with $V\approx0$ grows with $N$ and $\gamma$, and both numbers are much less than unity. The eigenvalues quantify the second derivatives of the potential and thus the mass-squared of the corresponding degree of freedom. The first derivatives necessarily vanish at an extremum, so to lowest order this information is sufficient to roughly quantify the possible quintessence potentials.   A small hierarchy in the masses is actually inevitable -- recall that the overall likelihood vanishes if any two eigenvalues are exactly equal. However, this raises the intriguing possibility of a quintessence scenario with two (or even three) ``active'' fields, but with masses such that the longer term behaviour will be dominated by a single field.

Anthropic approaches to understanding the value of the (apparent) cosmological constant assume that values of $\Lambda$ in the landscape are  dense enough for it to be likely that it supports the observed value of $\Lambda \sim 10^{-122}$; roughly speaking this requires that there are at least $1/\Lambda$ minima with positive vacuum energy within the total volume of fieldspace. The swampland constraint on $N$ and $\gamma$ sketched above rules out parameter values where the total volume of fieldspace is expected to admit any positive mimima. Conversely, the anthropically allowed subset of parameter space excludes not just the region compatible with the swampland hypothesis, but also any value of $\gamma$ (for fixed $N$) which does not admit large  numbers of positive minima.

A further consequence of this is that if we are indeed living in a universe with a positive cosmological constant, this singular  ``data point'' would constrain the overall parameter space of Gaussian landscape cosmologies. It is certainly true that any claim that our Universe is embedded in a generic multiverse is unfalsifiable. However, the scenario examined here provides an example of a specific  multiverse model whose parameter space can be meaningfully constrained by the mere fact of our existence as observers. In particular, for the very specific case of a Gaussian random field with a Gaussian power spectrum, the expected number of positive minima deceases super-exponentially with $N$. Consequently, these scenarios will provide food for thought for those who would seek to claim that {\em any \/} discussion of multiverse cosmology lies beyond the realm of empirical science.

 \acknowledgments We acknowledge useful discussions with  Mateja Gosenca,  Ali Masoumi, Jens Niemeyer, and Ben Schlaer. This work was supported in part by the Foundational Questions Institute (FQXi)  and the Silicon Valley Community Fund.

\appendix

\section{Proof that $0 < \sigma_1^2 < \sigma_0\sigma_2$} \label{Proof}

Consider the expression:

\begin{equation}
\int (k^2 - \frac{\sigma_1^2}{\sigma_0^2})^2 P(k) dk
\end{equation}

\noindent We must have that this expression is greater than 0, because the integrand is positive. Expanding the square leads to

\begin{equation}
\begin{split}
\int k^4 P(k) dk &- 2 \int k^2 \frac{\sigma_1^2}{\sigma_0^2}k^2 P(k) dk + \int \frac{\sigma_1^4}{\sigma_0^4} P(k) dk \\
&= \sigma_2^2 - 2 \frac{\sigma_1^4}{\sigma_0^2} + \frac{\sigma_1^4}{\sigma_0^2} \\
&= \sigma_2^2 - \frac{\sigma_1^4}{\sigma_0^2} > 0
\end{split}
\end{equation}
where the second line follows from the definition of the $\sigma$'s.

\end{document}